\documentclass[intlimits,twoside,a4paper]{article}

\usepackage[cp1251]{inputenc}

\usepackage[eqsecnum]{cmpj3}

\issue{2024}{27}{4}{43602}
\doinumber{10.5488/CMP.27.43602}

\title[Effect of arsenic doping on structural and electronic properties of MoSe$_2$ monolayer: an ab~initio study]
{Effect of arsenic doping on structural and electronic properties of MoSe$_2$ monolayer: an ab~initio study}
\author[B. Bradji, M. L. Benkhedir]{B. Bradji, 
	M. L. Benkhedir\orcid{0000-0001-8375-0998}\thanks{Corresponding author:\email{benkhedir@gmail.com}.}}
\address{
	Laboratory of Theoretical and Applied Physics (LPAT), Echahid Cheick Larbi Tebessi University, 12000 Tebessa, Algeria}


\date{Received July 2, 2024, in final form September 19, 2024}

\begin{document}
	
 \maketitle
	
\begin{abstract}	
In this paper, we studied the structural and electronic properties of MoSe$_2$ monolayer in its pure and doped forms, using the density functional theory (DFT), and the calculations were performed using Quantum Espresso (QE) software package. The doped systems are a MoSe$_2$ monolayer with a vacancy in Mo site (Mo vacancy system), the MoSe$_2$ monolayer with an As atom as substitutional for Mo atom (As(Mo) doped system) and an MoSe$_2$ monolayer with As atom in an interstitial site in the hollow location of the center of one ring of the structure between the plane Mo atoms and the plane containing Se atoms (As interstitial system). We calculated the formation energy of  various structures studied in Se-rich condition. We found that the As(Mo) doped system is a favorable configuration, whereas the As interstitial system is metastable. Different defects introduce midgap levels, which were interpreted according to the orbitals involved in their formation using the analysis of the band structure and DOS and PDOS of each system. The energy gap increases in all system structures and its value ranged between 1.5 eV and 1.73 eV, the Fermi level shifts toward the valence band for the Mo vacancy system, and As(Mo) doped system which suggests that it can be a $p$-type semiconductor, whereas Fermi level shifts to the conduction band for As interstitial system and suggests a $n$-type semiconductor behavior. The obtained results enable us to predict the possibility of using these systems in many applications, since it can be used in the As(Mo) doped system in photocatalysis or in photovoltaic applications in the visible light, and As interstitial system can be used in electronics applications in the infrared field.
%
%
\printkeywords
\end{abstract}

\section{Introduction}


After Novoselov et al. won the Nobel prize in 2010 for their successful preparation of graphene by exfoliation in 2004~\cite{neto2009electronic}, graphene and layered materials received increasing interest from researchers for their interesting structure and electronic properties \cite{haldar2015systematic, paul2017computational}. Graphene has very high carrier mobility at room temperature $(1.5\times10^{4}$ cm$^{2}$/Vs) and resistivity as low as $10^{-6}$ $\Omega$ which is much lower than copper and silver~\cite{bolotin2008temperature}, although graphene has a zero band gap which limits its possible applications in electronic and optoelectronic devices. It also has a low volumetric energy density and a large initial irreversible capacity that limit its application in lithium-ion batteries \cite{luo2022amorphous}. These limitations in possible graphene applications led to an intensive research work on the other layered 2D materials. Among these materials are the 2D transition metal dichalcogenides (TMDCs) that have fascinating properties, such as their intrinsic bandgap, highly efficient photovoltaic response, and tunable Seebeck coefficient~\cite{ma2020thermal}.
 The chemical formula of these materials is given by $MX_2$ ($M$ is a transition metal and $X$ is a chalcogen), and they have a honeycomb structure that is similar to that of graphene when seen from above \cite{menezes2021unveiling}, although the $M$ atoms form a layer sandwiched between two layers of $X$ atoms. The TMDCs can intrinsically exhibit a range of electronic properties, including semiconducting like in MoSe$_2$, MoS$_2$, WSe$_2$, and WS$_2$ or metallic behavior like in TiS$_2$ and VSe$_2$ or even supraconductive behavior like in TaS$_2$ and NbS$_2$~\cite{nair2008fine}. This property makes them important for applications in electronics and optoelectronics.
It is known that materials in nature are not devoid of defects of various kinds. These defects have a great advantage in improving the electrical, magnetic, and optical properties of matter. Therefore, point defects in particular have recently received great attention where Haldar and coworkers conducted a thorough study of the structural, electronic, and optical properties of many point defects in TMDCs which include single and double vacancies~\cite{haldar2015systematic}, while Ko\'os and his colleagues found that the chalcogen vacancy is  the most possible point defect that plays an important role in determining the optical properties in MoSe$_2$ \cite{koos2019influence}. By contrast, the vacancies Mo site in MoSe$_2$ was studied by Gao et al.~\cite{gao2018dual}, and the electrical, magnetic and optical properties of defect complexes involving a Mo vacancy and absorbed C, O, Si, Ge, and several transition metal atoms are also studied~\cite{haldar2015systematic}.
It is also very important to mention here that the band gap of TMDCs can be tuned by factors such as layer thickness, strain, or chemical doping~\cite{Kolobov2016}. This tunability makes them very important candidates for electronic device applications. It is important to mention here that in the case of chemical doping, there exist two types of doping, substitutional doping and metastable doping (like interstitial). Metastable doping can be voluntary or involuntary which is caused by the environment  like the substrate, metals used as contacts, humidity, etc.~\cite{zhang2019doping}. 
MoSe$_2$ has gained attention in scientific research and various technological applications due to its interesting properties~\cite{ayesh2022h2s}. It exists in nature in three phases. The first one is the trigonal prismatic 2H, the second is the trigonal 3R and the third is the octahedral phase 1T. In the octahedral phase, the MoSe$_2$ has a metallic behavior while in the two first phases, it has a semiconductive behavior~\cite{khan2021emerging}. It was also reported that physical and electronic properties of MoSe$_2$ can be changed by doping with different elements like the transition metals or chalcogens or other elements. Benquan et al. reported that MoSe$_2$ doped with Ta can be used in the detection of toxic gases~\cite{liang2022gas}. In this study, Ta atom replaced Se atom. Dongzhi et al. were able to synthesize Pd-decorated MoSe$_2$ used as a sensor of NH$_3$~\cite{zhang2019fabrication}. Yang et al. synthesized a MoSe$_2$ nanosheet array with layered MoS$_2$ heterostructures with high performance in hydrogen evolution and ions storage \cite{yang2017mose2}. Jianlin et al. studied the effect of nonmetal doping (N, C, and Si) of MoSe$_2$ on Schottky regulation of graphene/MoSe$_2$~heterojunction~\cite{he2023theoretical}. Yafei et al. studied the effect of nonmetal atom doping of MoSe$_2$ on its electronic properties and its possible application like in photocatalyse \cite{zhao2017first}, where they showed that nonmetal dopants with an odd number of valence electrons doped MoSe$_2$ have $n$ or $p$-type conductivity and better photocatalytic efficiency.
In literature there is an important number of studies on doping MoSe$_2$ with different atoms, although doping with an arsenic (As) atom is almost alike, and this encouraged us to contribute to the study of the effect of As doping.
 In this work, we conduct a theoretical study using the DFT for the effect of the As atom on the electronic properties, since we consider the As atom  as a substitute atom in the Mo site and compare it with the case of the vacancy in the Mo atom site. To expand the study further we also studied As atom as an interstitial atom in MoSe$_2$ monolayer.

\section{Computational details}

\begin{figure*} 
    \centering
	\includegraphics[width=0.95\textwidth]{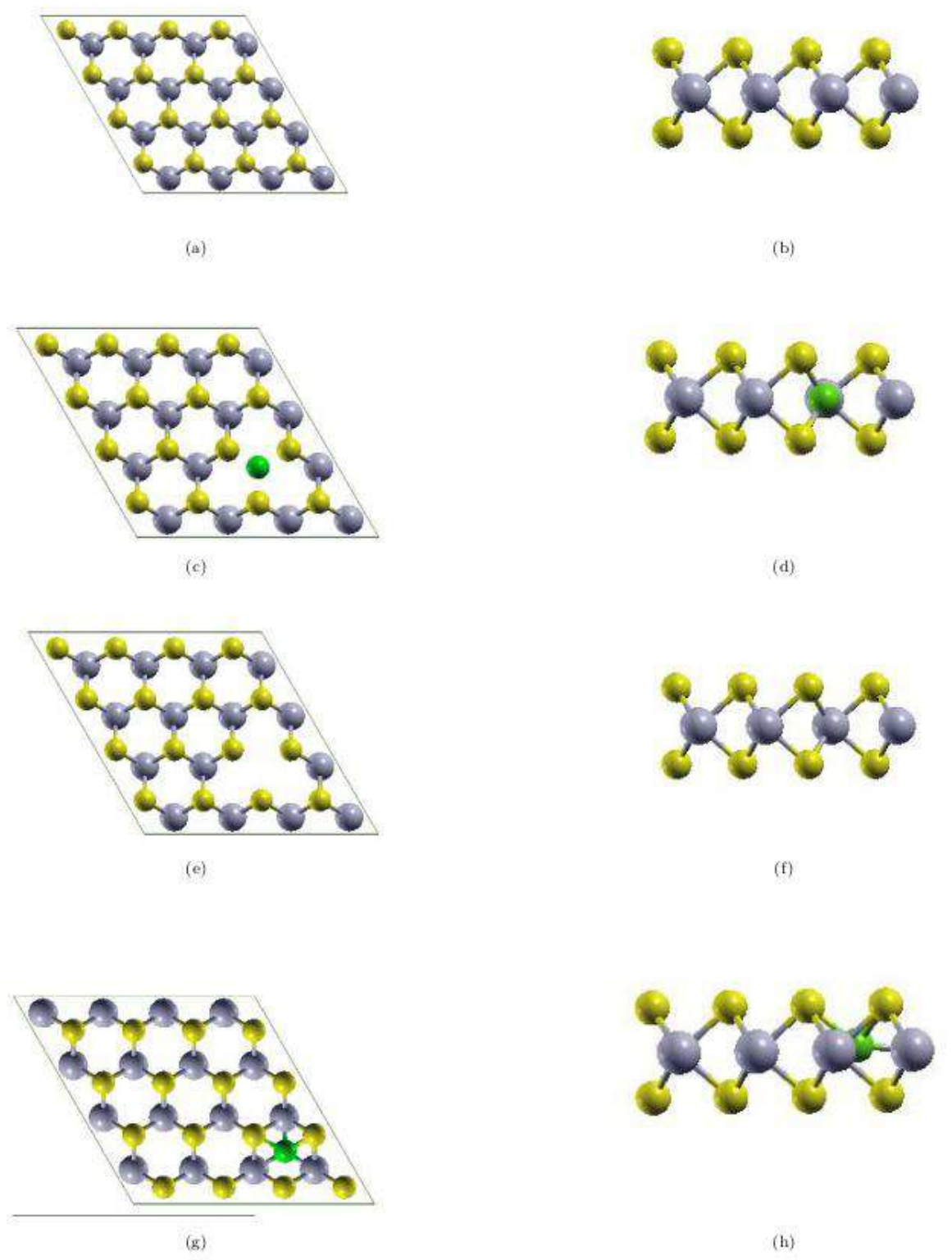}
	\caption{(Colour online) The crystal structure of $4\times4\times1$ supercell monolayer of MoSe$_2$ with and without defects, where (a) and (b) top and side views of pure MoSe$_2$ structure respectively, (c) and (d) top and side views MoSe$_2$ with As(Mo) doped structure respectively, (e) and (f) top and side views of  MoSe$_2$  with Mo vacancy respectively, (g) and (f) top and side views of  MoSe$_2$ with As interstitial respectively. The gray, yellow and green spheres denote the Mo, Se and As atoms, respectively.}
	\label{fig:mesh1}
\end {figure*}

 In this work, all calculations are done using density functional theory (DFT) based on the (PBE-GGA) approximation, whereas the methods of planar wave (PAW) and ultra-soft (US) atomic pseudopotentials (PPs) have been employed to calculate electronic properties \cite{giannozzi2009quantum, perdew1996generalized}, which were extracted from Standard Solid State PPs (SSSP). All the calculations that we have done are approached using the Quantum Espresso (QE) software package \cite{giannozzi2009quantum}. We built the supercell by repeating a unit cell along the $x-y$ plane by $4\times4\times1$ with a  15~\AA~ vacuum created along the $z$-axis i.e., in an out-of-plane direction to avoid the false interaction with replicas. We get a system that has 48 atoms as shown in figure~\ref{fig:mesh1}, of which 36 are Se atoms and 16 are Mo atoms. So the defect coefficient will be $6,25\%$  with respect to the number of Mo atoms when we replace the Mo with an As atom. All parameters used have been chosen with convergence in mind, where the parameter $a=3.32$~\AA~was chosen for a single cell. The first Brillouin zone was employed with a Monkhorst-Pack $K$-point mesh of $2\times2\times1$ for the SCF calculation, while more dense $k$-points were used for calculation in DOS and PDOS, which were estimated at $8\times8\times1$ \cite{monkhorst1976special}, and a plane-wave cut-off of 50~Ry, the convergence threshold on forces and estimated SCF accuracy were taken at $2\times10^{-3}$ Ry and $10^{-6}$, respectively, and the Murnaghan state equation was also used to calculate the value of the minimum energy \cite{murnaghan1944compressibility}.

\section{Formation energy}

The formation energies for different defects of monolayer MoSe$_2$ were calculated as:

\begin{equation}\label{eq:first}
	E_f=E_{\text{def}}-E_{\text{pure}}-\sum_{i}n_i\mu_i
\end{equation}

\begin{table*}[!b]
	
	\caption{Chemical potentials of Mo, Se and As atoms in (Ry).}
	\vspace{3mm}
	\centering
	\begin{tabular}{|c|c|c|}
		\hline
		\textbf   &  \textbf Chemical potential (Ry) \\
		\hline
		$\mu_{\text{Mo}}$  & $-139.746$   \\
		\hline
		$\mu_{\text{Se}}$    &    $-20.777$   \\
		\hline
		$\mu_{\text{As}}$       &  $-20.456$  \\
		\hline
	\end{tabular}
	\label{tab:Table.1}
\end{table*}

\begin{table*}[!b]
	\caption{Formation energies for the various cases studied in (eV).}
	\vspace{3mm}
	\centering
	\begin{tabular}{|c|c|c|}
		\hline
		\textbf{structure}   &  \textbf{$E_f$} (eV)  \\
		\hline
		MoSe$_2$  with Mo vacancy & 2.670    \\
		\hline
		MoSe$_2$  with As(Mo) doped     &   1.839   \\
		\hline
		MoSe$_2$  with As interstitial        &   17.422  \\
		\hline
	\end{tabular}
	
	\label{tab:Table.2}
\end{table*}

\begin{table*}[!b]
	\caption{Energy gap of different structures in (eV).}
	\vspace{3mm}
	\centering
	\begin{tabular}{|c  |l|}   
		\hline
		\textbf{structure }                      &  \textbf{$E_g$}(eV) \\
		\hline
		Pure MoSe$_2$  & 1.41 \\
		\hline
		MoSe$_2$ with Mo vacancy     & 1.66             \\
		\hline 
		MoSe$_2$ with As(Mo) doped    & 1.73                            \\
		\hline
		MoSe$_2$ with As interstitial   & 1.5                           \\
		\hline
	\end{tabular}
	\label{tab:Table.3}
\end{table*}

Where $E_{\text{def}}$ represents the total energy of a supercell containing a single defect, $E_{\text{pure}}$ is the total energy of a pristine MoSe$_2$  supercell with the same size, $\mu_i$ is the chemical potential and $n_i$ is the number of atoms added or removed of species $i$ ($n_i> 0$ added and $n_i<0$ removed)~\cite{menezes2021unveiling,haldar2015systematic,prucnal2021chlorine}.
 For the calculation of  chemical potentials, we use the bulk bcc phase of Mo and Se, and the bulk rhombohedral phase of As.
The experimental growth conditions mainly control the values of the chemical potentials, we consider the Se-rich conditions, where both $\mu_{\text{Mo}}$  and $\mu_{\text{Se}}$ are linked by this thermodynamic relationship $\mu_{\text{MoSe}_2} = \mu_{\text{Mo}} +2\mu_{\text{Se}}$, where $\mu_{\text{MoSe}_2}$ represents the $E_{\text{tot}}$ of the pristine unite cell and the values of chemical potentials are presented in table~\ref{tab:Table.1}.
The formation energies in Se-rich conditions of different structures studied are presented in table~\ref{tab:Table.2}. It can be observed that the MoSe$_2$ doped with As atom in a substitutional cite of Mo is a favorable configuration looking at its small formation energy 1.839 eV. The substitutional site of Se is also a stable configuration and the case was studied in \cite{zhao2017first}. The formation energy of As interstitial defect is 17.422 eV which suggests that the formation of this defect is possible but it looks like a metastable one.

The values of the energy gap of all studied structures are shown in table~\ref{tab:Table.2}.

\section{Result and discussion}
	
\begin{figure*}
		\centering
		\includegraphics [width=0.9\textwidth]{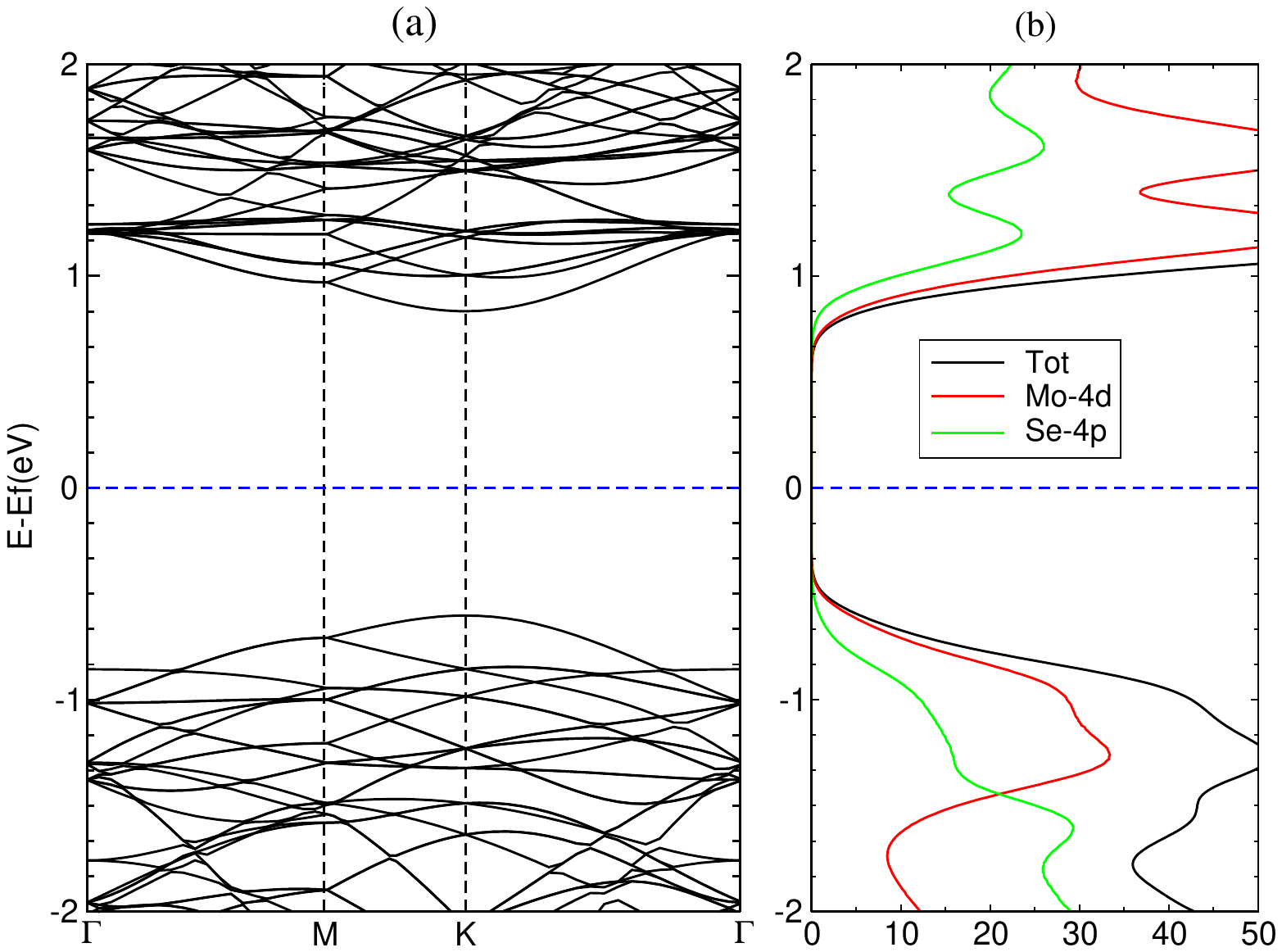}
		\caption{(Colour online) (a) Band structure of pure MoSe$_2$ monolayer, (b) DOS and PDOS of pure MoSe$_2$ monolayer.}
		\label{fig:mesh2}
	\end{figure*}

\begin{figure*}
		\centering
		\includegraphics[width=0.9\textwidth]{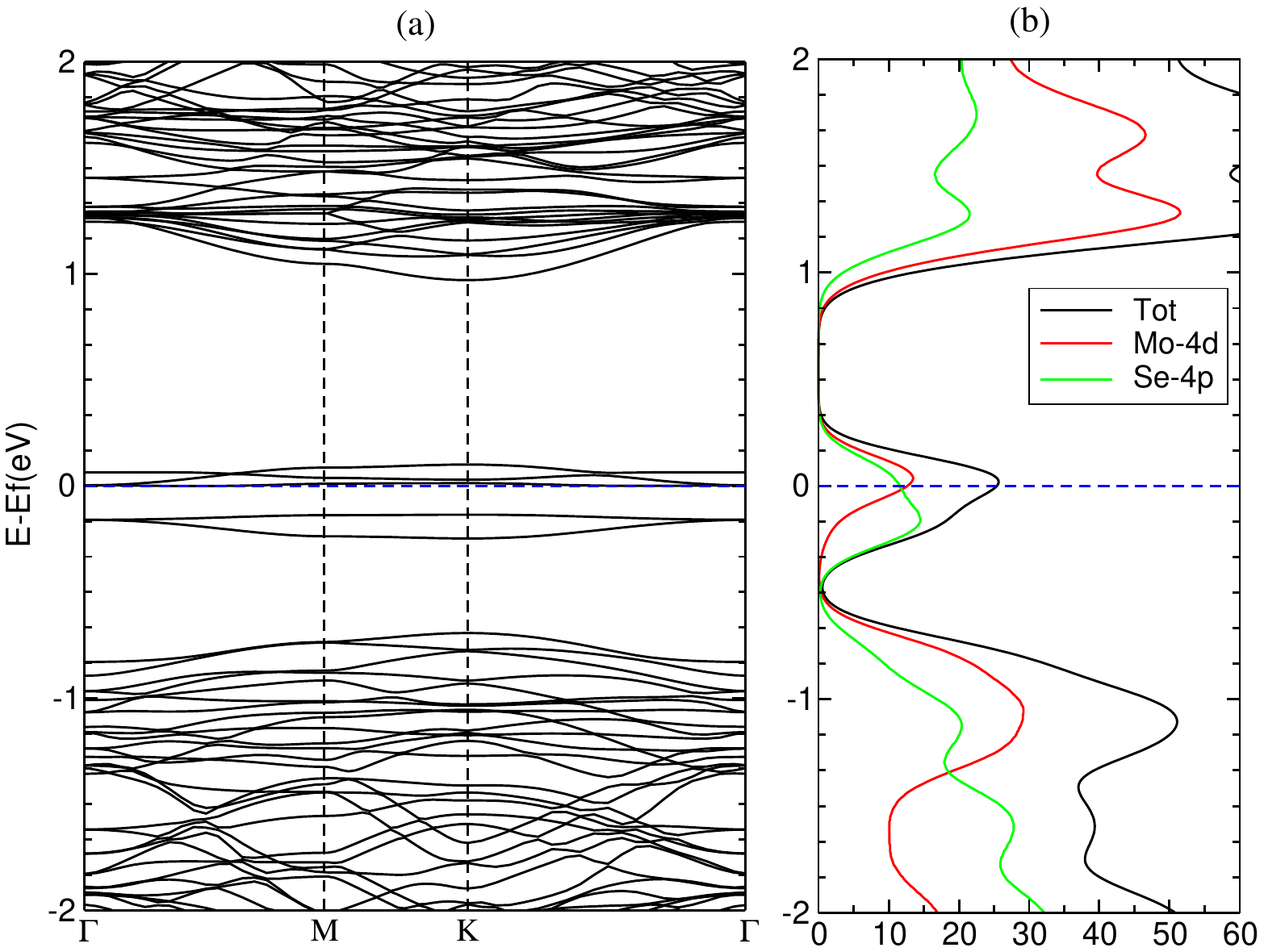}
		\caption{(Colour online) (a) Band structure of MoSe$_2$ monolayer with Mo vacancy, (b) DOS and  PDOS of MoSe$_2$ monolayer with Mo vacancy.}
		\label{fig:mesh3}
	\end{figure*}

In figure~\ref{fig:mesh2} we present the band structure and total and projected density of state (DOS and PDOS) of pure MoSe$_2$ monolayer.  The results presented in  figure~\ref{fig:mesh2} show that MoSe$_2$  monolayer is a semiconductor with a direct bandgap at the $K$ point of the Brillouin zone. The position of Fermi level suggests that this material is  an $n$-type semiconductor.  Since we are interested in the DOS in the vicinity of the bandgap, we have presented it between $-2$~eV and $+2$~eV.  The DOS at the top of the valence band and the bottom of the conduction band are mostly constituted of Mo $4d$ and Se $4p$ orbitals. These results are in good agreement with other studies \cite{vinturaj2023theoretical,kumar2012electronic,ramasubramaniam2012large,zhang2014direct}.
As it can be seen in figure~\ref{fig:mesh3}, the introduction of a vacancy in a Mo atom site increases the bandgap of the system and creates two different localized defect levels at the middle of the bandgap where the first can be attributed to the non-contributing  Mo $4d$ orbital in bonding (vacancy) at $+0.05$~eV, while the second can be attributed to the non-contributing Se $4p$ orbital in bonding at $-0.3$~eV. The position of  Fermi level in the system with a vacancy suggests that it is a $p$-type semiconductor. These  results are in good agreement with previous studies \cite{vinturaj2023theoretical}, \cite{zhao2021high}. 
	
	\begin{figure*}
		\centering
		\includegraphics[width=0.9\textwidth]{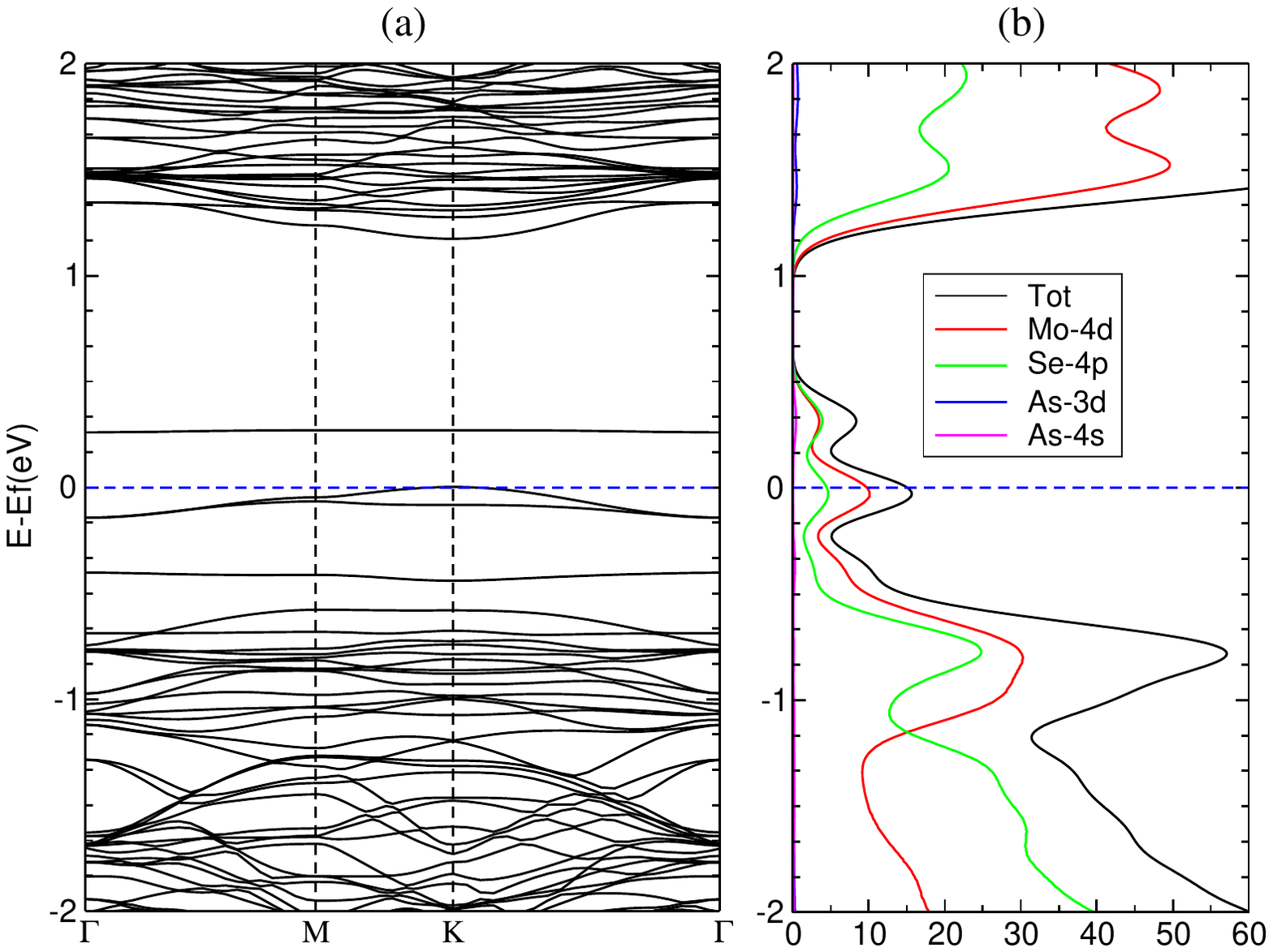}
		\caption{(Colour online) (a) Band structure of MoSe$_2$ monolayer with substutional As in Mo cite, (b) DOS and PDOS of MoSe$_2$ monolayer with As(Mo) doped structure.}
		\label{fig:mesh4}
	\end{figure*}

	\begin{figure*}
		\centering
		\includegraphics[width=0.9\textwidth]{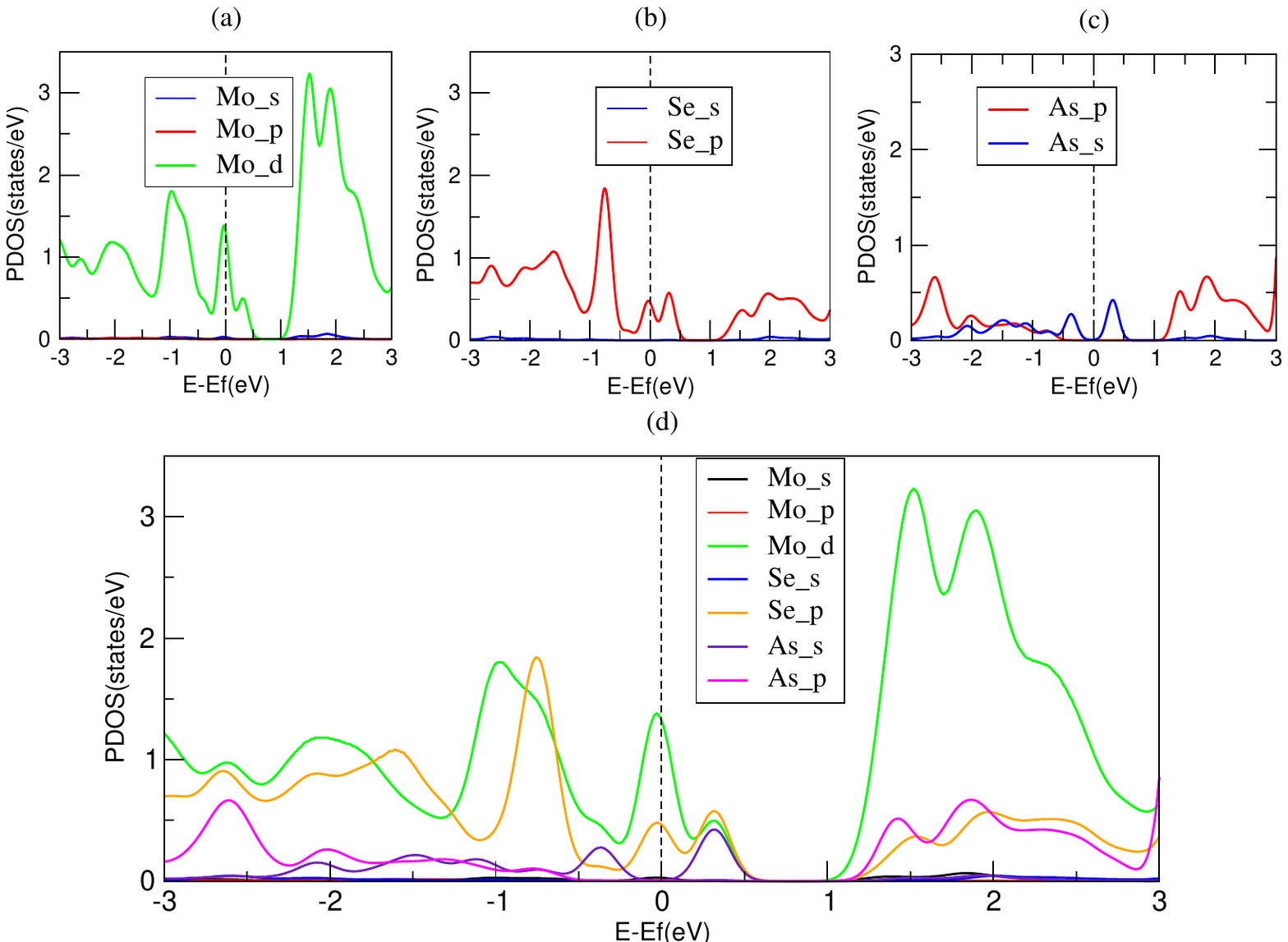}
		\caption{(Colour online) PDOS in As(Mo) doped system of: (a) per atom for Mo close to the As position, (b) per atom of Se that occupy positions  close to the As atom position, (c) As substitutional atom, (d) all previous cases for easy comparison.}
		\label{fig:mesh5}
	\end{figure*}
	
	\begin{figure*}
		\centering
		\includegraphics[width=0.9\textwidth]{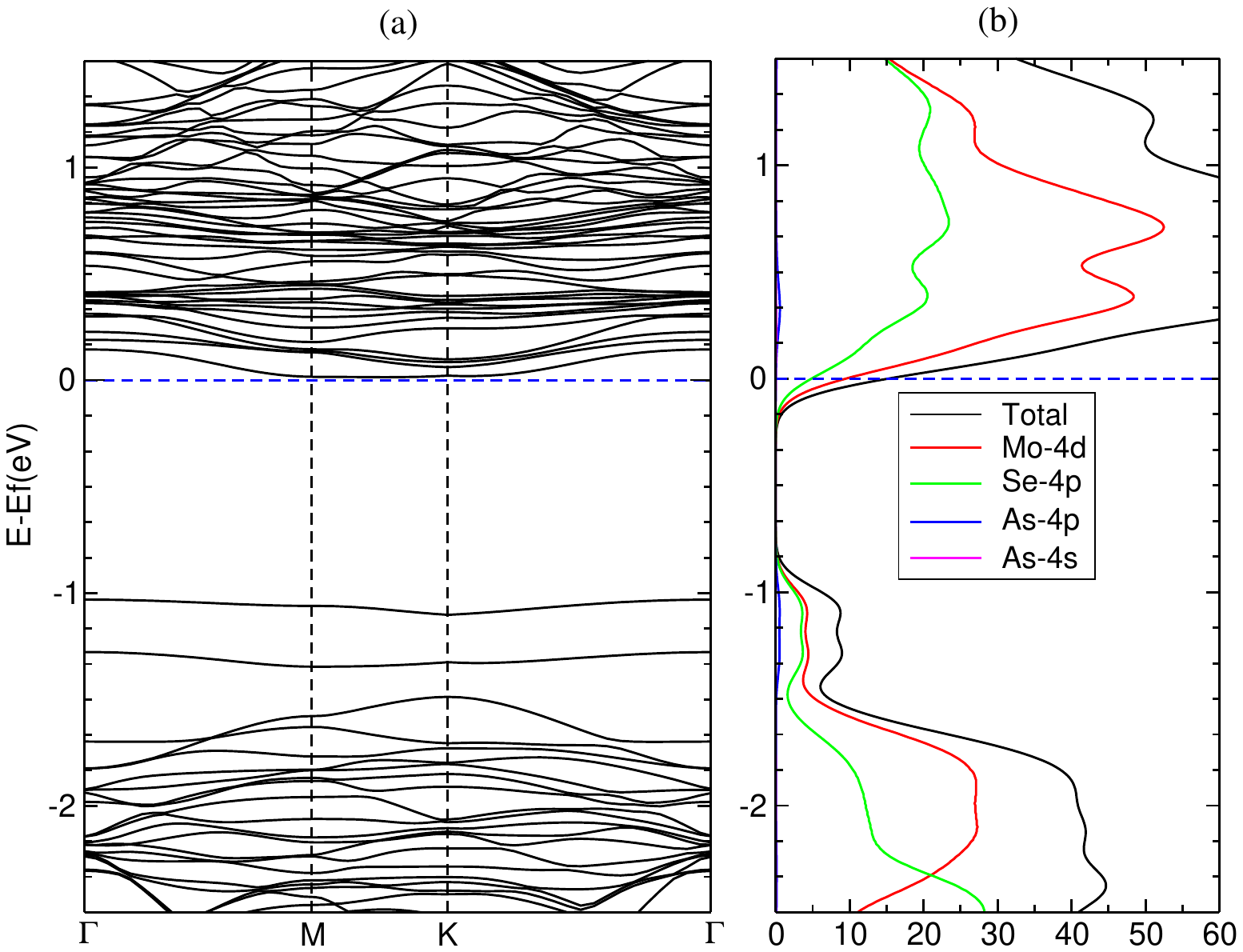}
		\caption{(Colour online) (a) Band structure of MoSe$_2$ monolayer with As interstitial, (b) DOS and PDOS of MoSe$_2$ monolayer with As interstitial.}
		\label{fig:mesh6}
	\end{figure*}
	
	\begin{figure*}
		\centering
		\includegraphics[width=0.9\textwidth]{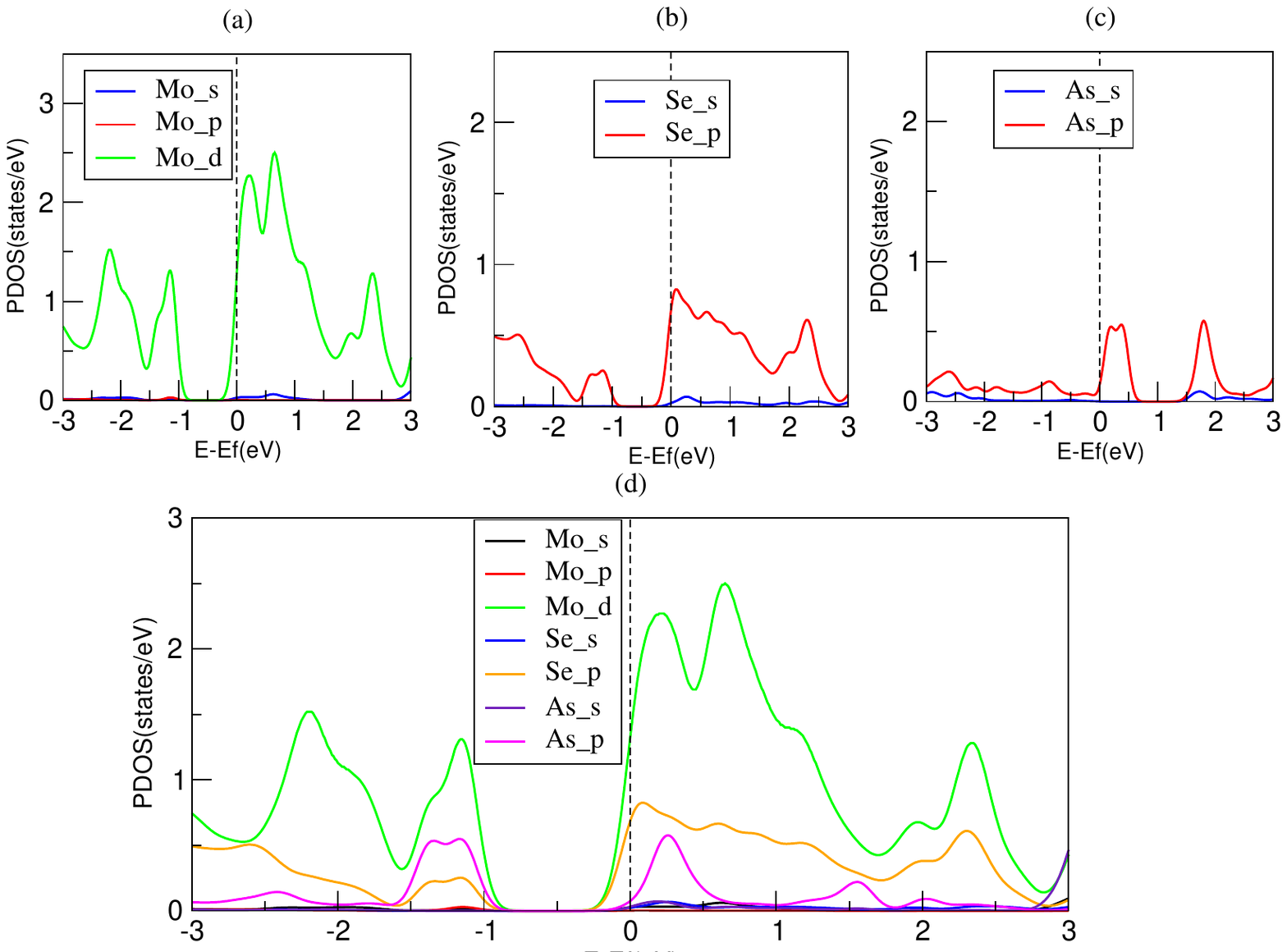}
		\caption{(Colour online) PDOS in interstitial As system of: (a) per atom for Mo around the interstitial As, (b) Se  around the interstitial As, (c) As interstitial atom and (d) all previous cases for easy comparison.}
		\label{fig:mesh7}
	\end{figure*}
	
Figure~\ref{fig:mesh4} shows the band structure and the total and projected DOS of a monolayer MoSe$_2$ containing a substitutional As atom in a Mo vacancy site. The bandgap of the As(Mo) doped system increases as compared to the pristine system and the system containing a vacancy as shown in table~\ref{tab:Table.3}, the DOS and PDOS show that the top and the bottom  of the valence band and the  conduction band respectively are composed of Mo $4d$ and Se $4p$ orbitals. The As(Mo) system shows a pair of defect levels in the bandgap, the first and the highest one is at 0~eV (Fermi level), while the second is slightly lower and is at $+0.35$~eV and a shoulder at the tail of the valence band at about $-0.35$~eV as it is obviously seen in figure~\ref{fig:mesh4}b. 
The PDOS in figure~\ref{fig:mesh4} does not help to understand the origin of these defect levels, where we cannot see the contribution of the As atom because the PDOS contains the contribution of all Mo and Se atoms  (15 and 32 atoms, respectively), since we have only one atom of As, that is why we have drawn the PDOS of per atom for Mo and Se that occupy positions close to the As atom position and the PDOS for As atom itself (as seen in  figure~\ref{fig:mesh5}). That leads us to attribute the defect level at 0~eV to Mo $4d$ and Se $4p$ orbitals like in vacancy defect, while the defect at $+0.35$~eV can be attributed to Se $4p$ and Mo $4d$ and As $4s$ orbitals,  the defect at $-0.35$~eV can be attributed to the As $4s$ and Se $4p$ orbitals. This shows us that when we introduce an As atom in substitutional site of a Mo atom, the bandgap increases. The Fermi level moves toward the valence band and three different defect levels appear in the midgap. The first is located at 0~eV, the second at $+0.35$~eV above Fermi level and the third at $-0.35$~eV below Fermi level. The volume of an As atom is smaller than the volume of Mo and Se atoms and this can explain that the DOS of an As (Mo) system is a combination of the effect of the As atom itself and the vacancy in the Mo site. The same observation is reported for the Sb(Mo) system in MoS$_2$ structure \cite{menezes2021unveiling,zhong2019electronic}. 
The widening of the bandgap and the introduction of new defect levels increases the optical absorption in the visible light spectral region. This enhances the possibilities of using the As(Mo) system in photocatalysis or in photovoltaic applications~\cite{zhao2017first,ayesh2022h2s}.
During the experimental preparation of As(Mo) system, one could suppose that the As atom occupies an interstitial site in the MoSe$_2$ monolayer. The same case could be envisaged when using the MoSe$_2$ monolayer in an environment containing As. For this we studied a MoSe$_2$ system containing an As atom in an interstitial site between the plane Mo atoms and the plane containing Se atoms as it is shown in figure~\ref{fig:mesh1}g and h. 
Figure~\ref{fig:mesh6}a and b show the band structure and total and projected DOS of a monolayer MoSe$_2$ containing interstitial As atom. We can see a slight increase in the bandgap compared  to the pristine system. The presence of an interstitial As in the MoSe$_2$ monolayer produces two single defect levels in the midgap, where they are located at $-1.04$~eV and $-1.29$~eV. 
To explain the reason for these two levels, it is necessary to draw the DOS and PDOS  of the interstitial As atom, as well as the DOS and PDOS of the Mo atom and As atom surrounding the As interstitial atom as shown in figure~\ref{fig:mesh7},
shows that the defect levels can be attributed to the Mo $4d$, Se $4p$, and As $4p$ orbitals. We also notice a very simple distribution of Mo $4d$ orbital for the defect level in $-1.04$~eV.
In this case, Fermi level shifted towards the conduction band, so the MoSe$_2$ monolayer can be doped $n$-type by As interstitial atom. This indicates the possibility of using this compound in  optoelectronics and electronics applications in the infrared field and photocatalysis in the visible light region \cite{zhao2017first}, \cite{balasubramaniam2019engineering}.
	
\section{Conclusion}

In conclusion, we have studied the structural and electronic properties of pure MoSe$_2$ monolayer and their doped forms represented by MoSe$_2$ monolayer with Mo vacancy, As(Mo) doped system, and with As interstitial atom. These structures are observed experimentally in MoS$_2$.
First, we calculate the formation energies in Se-rich condition. We found that the As(Mo) doped system is likely to be formed, whereas the As interstitial  system is unlikely to be formed and is metastable. Then, by studying the band structure and the PDOS of different systems and comparing them with the pure structure, we arrived at the existence of impurity levels in the band edge, which are estimated by two different levels for MoSe$_2$ with Mo vacancy attributed to the noncontribution of Mo $4d$ atom for the first level and the noncontributing of Se $4p$  atom for the other level, and by a pair of the defect level for As(Mo) doped system which attributed to $4p$  Se and $4d$ Mo and As $4s$.
 These results can explain that the contribution of orbitals in As(Mo) doped system is a mix between  the effect of the As atom and the vacancy effect due to  the small size of the As atom compared to the Mo and Se atoms.
   In the As interstitial system we can see two levels in the midgap attributed to the Mo $4d$ and Se $4p$ orbitals.
   Next, we found that the gap energy increases in all cases studied, whereas Fermi level position suggests that Mo vacancy system and As(Mo) system are $p$-type semiconductor and the As interstitial system is $n$-type semiconductor. These results enhance the possibility of using the As(Mo) doped system in photocatalysis and photovoltaic applications in visible light and using the As interstitial system in electronics applications and photocatalysis in the infrared and visible regions.  

\bibliographystyle{cmpj}
\bibliography{bradji}

\begin{thebibliography}{10}
\providecommand{\url}[1]{\texttt{#1}}
\providecommand{\urlprefix}{URL }
\expandafter\ifx\csname urlstyle\endcsname\relax
  \providecommand{\doi}[1]{doi:\discretionary{}{}{}#1}\else
  \providecommand{\doi}{doi:\discretionary{}{}{}\begingroup
  \urlstyle{rm}\Url}\fi
\providecommand{\eprint}[2][]{\url{#2}}

\bibitem{neto2009electronic}
Neto~A.~C., Guinea~F., Peres~N.~M., Novoselov~K.~S., Geim~A.~K., Rev. Mod.
  Phys., 2009, \textbf{81}, No.~1, 109, \doi{10.1103/RevModPhys.81.109}.

\bibitem{haldar2015systematic}
Haldar~S., Vovusha~H., Yadav~M.~K., Eriksson~O., Sanyal~B., Phys. Rev. B, 2015,
  \textbf{92}, No.~23, 235408, \doi{10.1103/PhysRevB.92.235408}.

\bibitem{paul2017computational}
Paul~J., Singh~A., Dong~Z., Zhuang~H., Revard~B., Rijal~B., Ashton~M.,
  Linscheid~A., Blonsky~M., Gluhovic~D., J. Phys.: Condens. Matter, 2017,
  \textbf{29}, No.~47, 473001, \doi{10.1088/1361-648X/aa9305}.

\bibitem{bolotin2008temperature}
Bolotin~K.~I., Sikes~K.~J., Hone~J., Stormer~H., Kim~P., Phys. Rev. Lett.,
  2008, \textbf{101}, No.~9, 096802, \doi{10.1103/PhysRevLett.101.096802}.

\bibitem{luo2022amorphous}
Luo~Y., Liu~Q., Yang~L., Yan~Y., Quim. Nova, 2022, \textbf{45}, 654--658,
  \doi{10.21577/0100-4042.20170864}.

\bibitem{ma2020thermal}
Ma~J.-J., Zheng~J.-J., Li~W.-D., Wang~D.-H., Wang~B.-T., Phys. Chem. Chem.
  Phys., 2020, \textbf{22}, No.~10, 5832--5838, \doi{10.1039/D0CP00047G}.

\bibitem{menezes2021unveiling}
Menezes~M.~G., Ullah~S., Phys. Rev. B, 2021, \textbf{104}, No.~12, 125438,
  \doi{10.1103/PhysRevB.104.125438}.

\bibitem{nair2008fine}
Nair~R.~R., Blake~P., Grigorenko~A.~N., Novoselov~K.~S., Booth~T.~J.,
  Stauber~T., Peres~N.~M., Geim~A.~K., Science, 2008, \textbf{320}, No. 5881,
  1308--1308, \doi{10.1126/science.1156965}.

\bibitem{koos2019influence}
Ko{\'o}s~A.~A., Vancs{\'o}~P., Szendro~M., Dobrik~G., Antognini~Silva~D.,
  Popov~Z.~I., Sorokin~P.~B., Henrard~L., Hwang~C., Bir{\'o}~L.~P.,
  Tapaszt{\'o}~L., J. Phys. Chem. C, 2019, \textbf{123}, No.~40, 24855--24864,
  \doi{10.1021/acs.jpcc.9b05921}.

\bibitem{gao2018dual}
Gao~D., Xia~B., Wang~Y., Xiao~W., Xi~P., Xue~D., Ding~J., Small, 2018,
  \textbf{14}, No.~14, 1704150, \doi{10.1002/smll.201704150}.

\bibitem{Kolobov2016}
Kolobov~A.~V., Tominaga~J., From 3D to 2D: Fabrication Methods, Springer
  International Publishing, Cham, 2016, 79--107,
  \doi{10.1007/978-3-319-31450-1_4}.

\bibitem{zhang2019doping}
Zhang~K., Robinson~J., MRS Adv., 2019, \textbf{4}, No. 51--52, 2743--2757,
  \doi{10.1557/adv.2019.391}.

\bibitem{ayesh2022h2s}
Ayesh~A.~I., Phys. Lett. A, 2022, \textbf{422}, 127798,
  \doi{10.1016/j.physleta.2021.127798}.

\bibitem{khan2021emerging}
Khan~Z.~H., Emerging trends in nanotechnology, Springer, 2021,
  \doi{10.1007/978-981-15-9904-0}.

\bibitem{liang2022gas}
Liang~B., Li~W., Ren~Q., Zhu~C., Li~J., Results Phys., 2022, \textbf{42},
  105978, \doi{10.1016/j.rinp.2022.105978}.

\bibitem{zhang2019fabrication}
Zhang~D., Li~Q., Li~P., Pang~M., Luo~Y., IEEE Electron Device Lett., 2019,
  \textbf{40}, No.~4, 616--619, \doi{10.1109/LED.2019.2901296}.

\bibitem{yang2017mose2}
Yang~J., Zhu~J., Xu~J., Zhang~C., Liu~T., ACS Appl. Mater. Interfaces, 2017,
  \textbf{9}, No.~51, 44550--44559, \doi{10.1021/acsami.7b15854}.

\bibitem{he2023theoretical}
He~J., Liu~G., Zhang~C., Wang~Y., Zhang~G., Micro Nanostruct., 2023,
  \textbf{180}, 207612, \doi{10.1016/j.micrna.2023.207612}.

\bibitem{zhao2017first}
Zhao~Y., Wang~W., Li~C., He~L., Sci. Rep., 2017, \textbf{7}, No.~1, 17088,
  \doi{10.1038/s41598-017-17423-w}.

\bibitem{giannozzi2009quantum}
Giannozzi~P., Baroni~S., Bonini~N., Calandra~M., Car~R., Cavazzoni~C.,
  Ceresoli~D., Chiarotti~G.~L., Cococcioni~M., Dabo~I., et~al., J. Phys.:
  Condens. Matter, 2009, \textbf{21}, No.~39, 395502,
  \doi{10.1088/0953-8984/21/39/395502}.

\bibitem{perdew1996generalized}
Perdew~J.~P., Burke~K., Ernzerhof~M., Phys. Rev. Lett., 1996, \textbf{77},
  3865--3868, \doi{10.1103/PhysRevLett.77.3865}.

\bibitem{monkhorst1976special}
Monkhorst~H.~J., Pack~J.~D., Phys. Rev. B, 1976, \textbf{13}, 5188--5192,
  \doi{10.1103/PhysRevB.13.5188}.

\bibitem{murnaghan1944compressibility}
Murnaghan~F.~D., PNAS, 1944, \textbf{30}, No.~9, 244--247,
  \doi{10.1073/pnas.30.9.244}.

\bibitem{prucnal2021chlorine}
Prucnal~S., Hashemi~A., Ghorbani-Asl~M., H{\"u}bner~R., Duan~J., Wei~Y.,
  Sharma~D., Zahn~D.~R., Ziegenr{\"u}cker~R., Kentsch~U., Krasheninnikov~A.~V.,
  Helm~M., Zhou~S., Nanoscale, 2021, \textbf{13}, No.~11, 5834--5846,
  \doi{10.1039/D0NR08935D}.

\bibitem{vinturaj2023theoretical}
Vinturaj~V., Yadav~A.~K., Jasil~T., Kiran~G., Singh~R., Singh~A.~K., Garg~V.,
  Pandey~S.~K., Bull. Mater. Sci., 2023, \textbf{46}, No.~3, 121,
  \doi{10.1007/s12034-023-02963-x}.

\bibitem{kumar2012electronic}
Kumar~A., Ahluwalia~P., Eur. Phys. J. B, 2012, \textbf{85}, 1--7,
  \doi{10.1140/epjb/e2012-30070-x}.

\bibitem{ramasubramaniam2012large}
Ramasubramaniam~A., Phys. Rev. B, 2012, \textbf{86}, 115409,
  \doi{10.1103/PhysRevB.86.115409}.

\bibitem{zhang2014direct}
Zhang~Y., Chang~T.-R., Zhou~B., Cui~Y.-T., Yan~H., Liu~Z., Schmitt~F., Lee~J.,
  Moore~R., Chen~Y., Lin~H., Jeng~H.-T., Mo~S.-K., Hussain~Z., Bansil~A.,
  Shen~Z.-X., Nat. Nanotechnol., 2014, \textbf{9}, No.~2, 111--115,
  \doi{10.1038/nnano.2013.277}.

\bibitem{zhao2021high}
Zhao~Y., Ren~Y., Coileain~C.~O., Li~J., Zhang~D., Arora~S.~K., Jiang~Z.,
  Wu~H.-C., Appl. Surf. Sci., 2021, \textbf{564}, 150399,
  \doi{10.1016/j.apsusc.2021.150399}.

\bibitem{zhong2019electronic}
Zhong~M., Shen~C., Huang~L., Deng~H.-X., Shen~G., Zheng~H., Wei~Z., Li~J., npj
  2D Mater. Appl., 2019, \textbf{3}, No.~1, 1, \doi{10.1038/s41699-018-0083-1}.

\bibitem{balasubramaniam2019engineering}
Balasubramaniam~B., Singh~N., Kar~P., Tyagi~A., Prakash~J., Gupta~R.~K., Mol.
  Syst. Des. Eng., 2019, \textbf{4}, 804--827, \doi{10.1039/C8ME00116B}.

\end{thebibliography}

\ukrainianpart

\title{Вплив легування миш'яком на структурні та електронні властивості моношару MoSe$_2$: першопринципні дослідження}
\author{Б. Браджі, 
	М. Л. Бенхедір\orcid{0000-0001-8375-0998}}
\address{Лабораторія теоретичної та прикладної фізики університету Ешахід Шейх Ларбі, 12000 Тебесса, Алжир}
\makeukrtitle

\begin{abstract}
	\tolerance=3000%
На основі теорії функціоналу густини досліджено структурні та електронні властивості моношару MoSe$_2$ у його чистій та легованій формах; розрахунки проводились за допомогою пакета програм Quantum Espresso. В якості легованих систем  вибирались моношар MoSe$_2$ із вакансією в околі Mo (система вакансій Mo), моношар MoSe$_2$ з атомом As як замісником атома Mo [легована система As(Mo)] та моношар MoSe$_2 $ з атомом As у міжвузловій ділянці в порожнистому місці поблизу центру одного кільця структури між площиною атомів Mo та площиною, що містить атоми Se (міжвузлова система As). Розраховано енергію утворення різних структур, досліджених у насиченому Se стані. Виявлено, що легована система As(Mo) є сприятливою конфігурацією, тоді як міжвузлова система As є метастабільною. Різні дефекти утворюють міжзонні щілини, які інтерпретувалися відповідно до орбіталей, залучених до їх формування, використовуючи аналіз зонної структури та DOS і PDOS кожної системи. Заборонена зона збільшується в усіх наявних структурах, а її значення міняється від 1,5 еВ до 1,73 еВ. Рівень Фермі зсувається в бік валентної зони для системи вакансій Мо та системи, легованої As(Mo), що свідчить про утворення напівпровідника $p$-типу, тоді як відповідне зміщення рівня Фермі в зону провідності для міжвузлової системи As характеризує поведінку напівпровідника типу $n$. Отримані результати дозволяють говорити про різноманітне прикладне використання таких систем, зокрема у фотокаталізі та фотоелектричних експериментах у видимому діапазоні світла [леговані системи As(Mo)], тоді як міжвузлову систему As можна використовувати в інфрачервоному діапазоні.
	\keywords перехідні дихалькогеніди металів, теорія функціоналу густини, електронна структура, легування
	
\end{abstract}
\lastpage
\end{document}